\documentclass[conference]{IEEEtran}
\usepackage{cite}
\usepackage{amsmath,amssymb,amsfonts}
\usepackage{algorithmic}
\usepackage{textcomp}
\usepackage{xcolor}
\usepackage{graphicx} %package to manage images
\graphicspath{ {./images/} }
\usepackage[rightcaption]{sidecap}
\usepackage{array}
\usepackage{wrapfig}
\usepackage[center]{caption}
\usepackage[utf8]{inputenc}
\usepackage{amsmath}
\usepackage{tikz}
\usepackage{tikzpagenodes}
\begin{document}

\title{Quantum Circuit Optimization of Arithmetic circuits using ZX Calculus}

\author{\IEEEauthorblockN{Aravind~Joshi}
\IEEEauthorblockA{\textit{Vellore Institute of Technology } \\
% \textit{Vellore Institute of Technology }\\
Chennai, India \\
% email address or ORCID
aravind.joshi2017@vitalum.ac.in}
\and
\IEEEauthorblockN{Akshara~Kairali}
% \IEEEauthorblockA{\textit{Electrical and Computer Engineering } \\
\IEEEauthorblockA{\textit{Vellore Institute of Technology } \\
% \textit{Vellore Institute of Technology }\\
Chennai, India \\
% email address or ORCID
akshara.kairali2020@vitalum.ac.in}
\and
\IEEEauthorblockN{Renju~Raju}
% \IEEEauthorblockA{\textit{Electrical and Computer Engineering } \\
\IEEEauthorblockA{\textit{Vellore Institute of Technology } \\
% \textit{Vellore Institute of Technology }\\
Chennai, India \\
% email address or ORCID
renjuraju.c2020@vitalum.ac.in}
\and
\IEEEauthorblockN{Adithya~Athreya}
\IEEEauthorblockA{\textit{Vellore Institute of Technology } \\
% \textit{Vellore Institute of Technology }\\
Chennai, India \\
% email address or ORCID
adithya.athreya2019@vitstudent.ac.in}
\and
\IEEEauthorblockN{Reena~Monica~P*}
% \IEEEauthorblockA{\textit{Electrical and Computer Engineering } \\
\IEEEauthorblockA{\textit{Vellore Institute of Technology } \\
% \textit{Vellore Institute of Technology }\\
Chennai, India \\
reenamonica@vit.ac.in}
\and
\IEEEauthorblockN{Sanjay~Vishwakarma}
% \IEEEauthorblockA{\textit{Quantum Department} \\
\IEEEauthorblockA{\textit{IBM Quantum} \\
% \textit{IBM Quantum}\\
San Jose, United States \\
sanjay.vishwakarma@ibm.com}
\and
\IEEEauthorblockN{Srinjoy~Ganguly}
% \IEEEauthorblockA{\textit{School of Technology \& Business} \\
\IEEEauthorblockA{\textit{Woxsen University} \\
% \textit{Woxsen University}\\
Hyderabad, India \\
srinjoy.ganguly@fractal.ai}
} 

\maketitle
\begin{tikzpicture}[remember picture,overlay,shift={(current page text area.south west)}]
    \node [above right]{\parbox{\textwidth}{*Corresponding Author}};
    \end{tikzpicture}
\begin{abstract}
Quantum computing is an emerging technology in which quantum mechanical properties are suitably utilized to perform certain compute-intensive operations faster than classical computers. Quantum algorithms are designed as a combination of quantum circuits that each require a large number of quantum gates, which is a challenge considering the limited number of qubit resources available in quantum computing systems. Our work proposes a technique to optimize quantum arithmetic algorithms by reducing the hardware resources and the number of qubits based on ZX calculus. We have utilised ZX calculus rewrite rules for the optimization of fault-tolerant quantum multiplier circuits where we are able to achieve a significant reduction in the number of ancilla bits and T-gates as compared to the originally required numbers to achieve fault-tolerance. Our work is the first step in the series of arithmetic circuit optimization using graphical rewrite tools and it paves the way for advancing the optimization of various complex quantum circuits and establishing the potential for new applications of the same.
\end{abstract}


%\section{Introduction}
%\IEEEPARstart Na\"ive Bayes\cite{shende2004minimal}, the simplest but most efficient Bayesian classifier, has been widely used for many decades \cite{lewis1998naive, domingos1997optimality}. A Bayesian classifier computes the probability for every possible label by the Bayes Theorem as shown in Definition \ref{def:1}.*/
\section{Introduction}
\IEEEPARstart{I}{n} classical computing, vast amounts of data cannot be computed simultaneously which makes it impractical to model and solve problems of such data-intensive nature. The large-scale fault-tolerant realization of quantum computers on the contrary, is promising due to its ability to compute these amounts of data simultaneously, opening up a wide range of applications like prime factorization of large numbers, molecular structuring to find new drugs and cybersecurity. Quantum circuit computations are performed using quantum algorithms, which make use of various arithmetic circuits such as subtraction, multiplication, and addition. Any implementation of these circuits is conditional to keeping the overall resources consumed at an acceptable level, which is achieved by the process of Quantum circuit optimization, usually at a gate level as a collection of gates generally contains all one-bit quantum gates and hence the 2-bit quantum ex-or gate or exclusive-or gate can also be represented as a combination of
\newpage 
these gates  \cite{barenco1995n}. Quantum circuit optimization
is a vital topic in the study of quantum circuits and can be defined as the transformation of given computations into novel circuits using fewer or simple gates while maintaining their functionality. Since the beginning, when quantum circuits were designed using quantum algorithms, research has been carried out for the compilation and optimization of these circuits. The techniques used target various facets like minimizing the qubits count \cite{beauregard2002circuit, shende2004minimal }, ancilla bits and garbage bits,reduction in the  limitations due to hardware resources  \cite{maslov2007linear} and in the number of gates that are more expensive when simulated in error-corrected harware \cite{paler2017fault}. For instance, the ancilla constant inputs and outputs are used for computation but are not useful since the input or output is garbage bit \cite{munoz2018quantum} and therefore the circuit overhead such as the ancilla and garbage outputs need to be minimized. 
An example of these optimization was the implementation of a reversible quantum integer multiplier that was a garbage output optimized circuit which achieved an efficiency of sixty to ninety per cent compared to the existing designs  \cite{jayashree2016ancilla}. Despite research in this field, the optimizations were costlier to implement in error-corrected hardware error  \cite{paler2017fault} and correction protocols were not investigated in the early work. When assessing optimal circuit constructions, these protocols place constraints on the cost metrics. Many papers have begun to recognize that to implement fault-tolerant circuits, Clifford + T universal gates can be utilized to overcome noise-error imposed limits. By avoiding uncontrollable errors caused by the quantum-bits interaction, also known as Quantum fault-tolerance, the circuits are more robust and accurate in design. The presented concept underlays the framework for large-scale quantum circuit optimization of fault-tolerant quantum circuits. The techniques like gate substitution, calculation of small (false-)normal forms for distinct families of circuits, and optimization of phase polynomials are the most common approaches for quantum circuit optimization  \cite{paler2017fault}. A novel method of quantum circuit optimization, formulated from the ZX-calculus is presented here. The circuits are first interpreted as ZX-diagrams, which provide a flexible, low-level language graphical description of quantum calculations \cite{duncan2020graph}. Using the rules of ZX-calculus, a strategy for simplification can be devised and to demonstrate that the final reduced diagram obtained can be translated back into a quantum circuit. This optimization method provides a new normal form which is desirable in size and reduces the T-gate count of quantum circuits. For Clifford + T gate circuits, this method helps us to investigate or explore gates that interfere with the Clifford architecture \cite{duncan2020graph}. The above is demonstrated by implementing a fault-tolerant 6-qubit multiplier, followed by the optimization of the same. The paper is organized as follows. In section II, the methodology behind the implementation of quantum gates is discussed, followed by a brief overview of ZX calculus. The section concludes by discussing the implementation of the Multiplier circuit. Section III highlights the key steps of the execution and optimization results. Section IV provides the conclusion to the work.

\section{Preliminaries}
Quantum gates act as building blocks to quantum circuits. They operate using qubits, also known as quantum bits, which have quantum information encoded in them. The entanglement property helps the quantum gate to avoid loss of information since the qubits are entangled inside a quantum gate. The Clifford gate sets are utilized to carry out the fault-tolerant circuit implementation. It is already shown that the GottesmanKnill theorem allows efficient classical simulation of
quantum circuits composed of gates in the Clifford set\cite{wecker2014liqui}. This makes it possible to simulate circuits with several gates in the order of 1000\cite{gottesman1998heisenberg, aaronson2004improved}. This however needs a quantum computer that needs to use gates outside of
the Clifford set so that the speed is maintained. It is commonly noted for most models to have Clifford group with an efficient set of generators while the non-clifford group gates require more expensive procedures to implement \cite{amy2013meet}.The most popular non-Clifford gate to use in conjunction with the  Clifford set is the T gate. 
 The gates included in the set, along with a few other elementary gates are described below.     

\subsection{Hadamard Gate}
\begin{figure}[h]
    \centering
    \includegraphics[width=2cm, height=1cm]{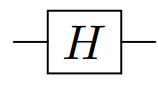}
    \caption{Hadamard Gate}
    \label{Fig 1}
\end{figure}

The Hadamard gate is used to create superposition of states in a single qubit. It is represented as shown in Fig \ref{Fig 1}
\newline
\newline
A superposition of $|$0 $>$ and $|$1 $>$ states can be visualized as the displacement away from the polar point of a Bloch sphere. The gate can be defined as follows –
\begin{equation}
H= (\frac{1}{\sqrt{2}})
\times \begin{bmatrix}
1 & 1\\
1 & -1
\end{bmatrix} 
\end{equation}\

\subsection{T Gate}
\begin{figure}[h]
    \centering
    \includegraphics[width=2cm, height=1cm]{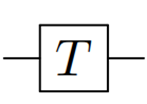}
    \caption{T Gate}
    \label{Fig 2}
\end{figure}

The T gate is used to induce a $\pi$/4 phase on a single qubit and can be related to the Phase gate and the fourth root of Pauli-Z. It is however not a Clifford gate. It is represented as shown in Fig \ref{Fig 2}
\newline
\newline
The gate can be defined as follows –
\begin{equation}
T= 
\begin{bmatrix}
1 & 0\\
0 & e^{(\frac{i\pi}{4})}
\end{bmatrix} 
\end{equation}

\subsection{Hermitian of T Gate}
\begin{figure}[h!]
    \centering
    \includegraphics[width=2cm, height=1cm]{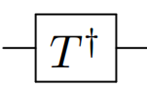}
    \caption{Hermitian of T Gate}
    \label{Fig 3}
\end{figure}

The Hermitian of T gate is used to induce a negative  $\pi$/4 phase on a single qubit and can be related to the S gate as well (as the product of the S gate with itself).
The  gate can be represented as shown in Fig \ref{Fig 3} –
\newline
\newline
\newline
\begin{equation}
{T}^{\intercal}= 
\begin{bmatrix}
1 & 0\\
0 & e^{(\frac{-i\pi}{4})}
\end{bmatrix} 
\end{equation}

\subsection{Phase Gate}
\begin{figure}[h]
    \centering
    \includegraphics[width=2cm, height=1cm]{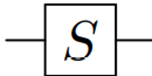}
    \caption{Phase Gate}
    \label{Fig 4}
\end{figure}

The Phase gate induces a 90-degree rotation about the Z-axis. It is represented as shown in Fig \ref{Fig 4}

%\newline
%\newline

The  gate can be defined as follows –
\newline
\begin{equation}
S= 
\begin{bmatrix}
1 & 0\\
0 & i
\end{bmatrix} 
\end{equation}

\subsection{Hermitian of Phase Gate}
\begin{figure}[h !]
    \centering
    \includegraphics[width=2cm, height=1cm]{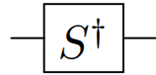}
    \caption{Hermitian of Phase Gate}
    \label{Fig 5}
\end{figure}

The Hermitian of Phase gate induces a 90-degree rotation about the Z-axis. It is represented as shown in Fig \ref{Fig 5}
\newline

The gate can be defined as follows –
\newline
\begin{equation}
{S}^{\intercal}= 
\begin{bmatrix}
1 & 0\\
0 & -i
\end{bmatrix} 
\end{equation}

\subsection{Not Gate}
\begin{figure}[h]
    \centering
    \includegraphics[width=2cm, height=1cm]{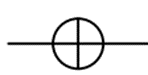}
    \caption{NOT Gate}
    \label{Fig 6}
\end{figure}

Similar to a classical NOT gate, the quantum NOT gate also reverses the state of a qubit. It is represented as shown in Fig \ref{Fig 6}
\newline
\newline
The gate can be defined as follows –
\newline
\begin{equation}
NOT= 
\begin{bmatrix}
0 & 1\\
1 & 0
\end{bmatrix} 
\end{equation}

\subsection{CNOT Gate}

\begin{figure}[h !]
    \centering
    \includegraphics[width=2cm, height=1.5cm]{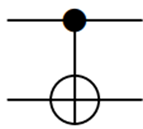}
    \caption{CNOT Gate}
    \label{Fig 7}
\end{figure}

The CNOT gate is a 2-qubit gate, where one qubit is a control qubit and the other is a target qubit, upon which the X gate is applied to if the control is true. It is represented as shown in Fig \ref{Fig 7}

The gate can be defined as follows -
\newline
\begin{equation}
CNOT= 
\begin{bmatrix}
1 & 0 & 0 & 0\\
0 & 1 & 0 & 0\\
0 & 0 & 0 & 1\\
0 & 0 & 1 & 0
\end{bmatrix} 
\end{equation}
\newline

\begin{figure}[h]
    \centering
    \includegraphics[width=8cm, height=1cm]{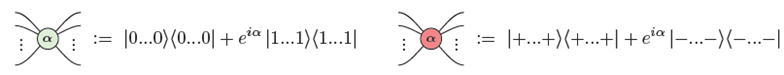}
    \caption{Z and X Spiders}
    \label{Fig 8}
\end{figure}

Having gone over the basic quantum gates and their function, the other concept that requires some background is ZX-Diagrams and their Calculus. ZX-diagrams are essentially a graphical representation of complex matrices of size ${2}^{n}  \times  {2}^{m}$. This notation allows calculations to be done with it, which can simplify the effort that it would otherwise take. A ZX-diagram consists of spiders and wires, which constitute the basic building blocks of ZX-Calculus. The thin lines are called wires, where the ones that are present on the left of the dot are inputs and any wires that are to the right side of the same are outputs \cite{duncan2020graph}. Spiders, also known as generators are linear maps with the probability of having many inputs or outputs. There are two types of spiders: the Z spider illustrated as a green dot and the X as a red dot where each represents a complementary set of bases. This can be seen in Fig \ref{Fig 8}. This linear mapping built using spiders is used to model different quantum gates. 

%\newline
%\newline

\begin{figure}[h]
    \centering
\includegraphics[width=0.15\textwidth]{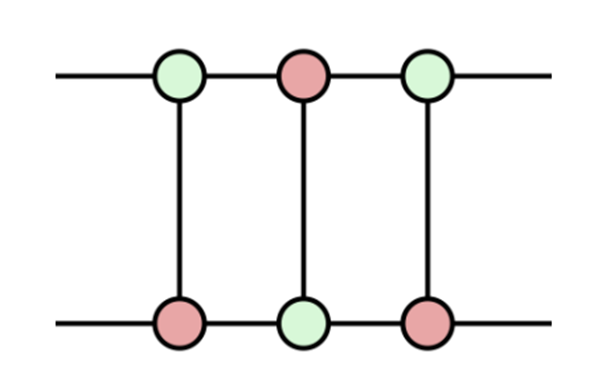}
    \caption{ZX-diagram of circuit with 3 CNOT Gates}
    \label{Fig 9}
\end{figure}

ZX-Calculus has been used in the simplification of circuits in the past as well. One popular method combines the `sum-of-stabilisers' method with an automated simplification strategy based on the ZX calculus. They adapted these techniques from the original setting of Clifford circuits with magic state injection to generic ZX-diagrams and show that, by interleaving this "chunked" decomposition with a ZX-calculus-based simplification strategy, stabiliser decompositions can be obtained that are many orders of magnitude smaller than existing approaches \cite{kissinger2022simulating}. Similarly, the usability of ZX-Calculus for small computations on quantum circuits and states is already discussed along with the Clifford computation and graphical proof of the Gottesman-Knill theorem, and the recent completeness theorems for the ZX-calculus that show that, in principle, all reasoning about quantum computation can be done using ZX-diagrams \cite{van2020zx}
There are two main differences between ZX diagrams and quantum circuits. ZX-diagrams need not conform to a rigid topological structure of circuits and hence can be deformed continuously. The second is that there are a set of rewrite rules for ZX-diagrams like the copy rule, Bi-algebra rule, and fusion rules collectively referred to as the ZX-calculus
An example of simplification of a ZX-diagram is demonstrated below -
%\newline
Let us begin with a ZX-diagram with 3 CNOT gates as shown in \ref{Fig 9}.
\newline
We can apply the bi-algebra rule to this.
\newline
\begin{figure}[h !]
    \centering
    \includegraphics[width=0.15\textwidth]{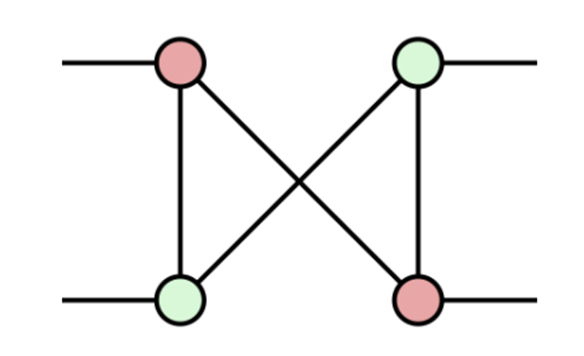}
    \caption{After Bi-algebra law is applied}
    \label{Fig 10}
\end{figure}

\begin{figure}[h]
    \centering
    \includegraphics[width=0.18\textwidth]{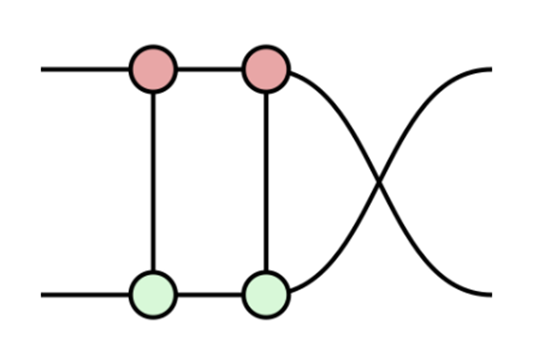}
    \caption{Circuit after swap gate}
    \label{Fig 11}
\end{figure}

Next, we can pull the gate through the swap as shown in Fig \ref{Fig 11}.
\newline
\newline
This is followed by merging the two red nodes and two green nodes.
\begin{figure}[h !]
    \centering
    \includegraphics[width=0.2\textwidth]{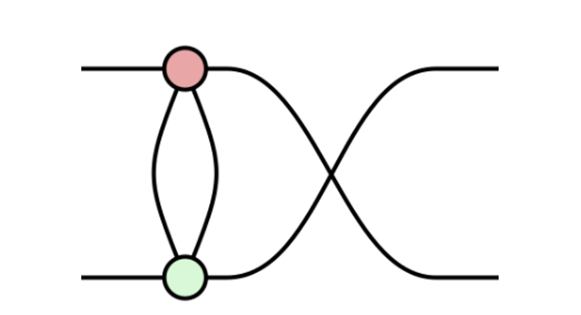}
    \caption{Merging the nodes}
    \label{Fig 12}
\end{figure}
To this, we can apply the Hopf law as shown in Fig \ref{Fig 13}.
\begin{figure}[h !]
    \centering
    \includegraphics[width=0.2\textwidth]{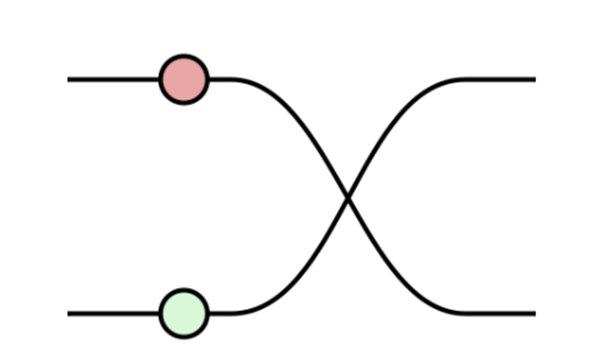}
    \caption{Circuit after Hopf law is applied.}
    \label{Fig 13}
\end{figure}
\newline
Finally, passing this through the swap gate removes the blank spider and gives us the simplified diagram as shown Fig \ref{Fig 14}.
\begin{figure}[h !]
    \centering
    \includegraphics[width=0.2\textwidth]{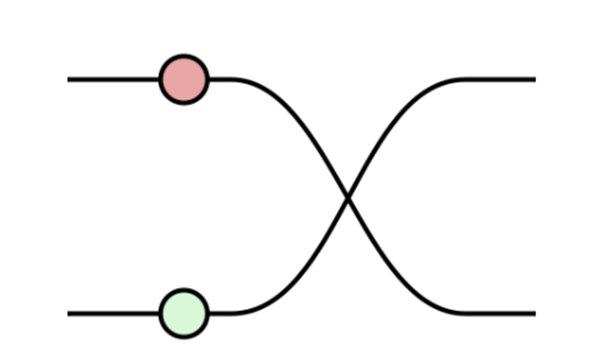}
    \caption{ZX-diagram of simplified circuit }
    \label{Fig 14}
\end{figure}

\section{Method}
 The fault tolerant implementation of quantum arithmetic circuits is gaining a lot of attention from the research community because of their fundamental use in a variety of applications and the need to adjust for physical quantum computers' proclivity for noise error. There are various such implementations, such as Integral multiplication, Integral Divider, Non-restoring square root. These form a class of fundamental arithmetic circuit implementations. For Instance, one implementation of the Integral Multiplier is by Thapliyal et al.,2019. The modules do not produce any garbage outputs and also restore inputs to their starting value and therefore is an efficient implementation in terms of Qubit use and T-gates. Other implementations include Galois Squaring and Exponentiation, Bilinear Interpolation and Reversible Square. All of these can be optimized for their T-count or T-depth in their fault-tolerant implementations as the quantum T gate, which is a non-Clifford gate, used to obtain a universal gate set.
\subsection{Integer Multiplier Circuit}    

For the implementation of Quantum Multiplier circuit using Shift and Add algorithm, conditional Adder circuits are utilized. In conditional adder circuit, two sets of input signals A and B are given, along with the ‘ctrl’ input. Quantum registers are used to store the following qubits, i.e., for a n bit conditional adder, a quantum register for first input A with n qubits and 2 qubits, a second quantum register for input B with n qubits and another quantum register with a single qubit for control input are used. Registers form the basis of quantum circuit. Quantum operations are forced sequentially on quantum bits to generate the result. The nth qubit in first Quantum Register is transformed to MSB and the n qubits in the second quantum register transforms to the other bits of the sum. All the other qubits are unchanged at the output. The circuit is as shown in Fig \ref{Fig 15}.
The conditional Adder circuits are placed in a shifted manner to perform multiplication. For the multiplication of two n bit numbers, n conditional adders are used. The first stage of adder can be replaced by an array of Toffoli gates, which helps to reduce the overall T-count in the circuit. The T gate count in a 6-bit conditional adder is 140, whereas Toffoli gate array has only 42 T gates.
There are 25 qubits in this quantum circuit, 6 of which correspond to the first input, 6 again to the second input and 13 ancilla inputs initialized to zeroes. Each of the 6 qubits in the second input acts as the control input to a corresponding conditional adder. At the end of computation, 12 ancilla bits contain the 12 product bits and the final ancilla input is unchanged.

\subsection{ZX Optimization}

ZX calculus is a graphical language which can be used to create graphical representations of quantum circuits. The calculus works by using Z and X functions, which allows us to modify the quantum circuit model while maintaining soundness of reasoning. By this way, the properties of circuits, protocols, and entanglement states, can be depicted in a visually clear and logically complete order.
ZX optimization begins with the ZX-diagrams being converted to graph-like form where every spider is only of the Z type which can be achieved using twisting and re-definition, and every spider is associated with some output and no non-zero phases. This is followed by the process of removing as many internal spiders in the graph-like form by pivoting and inverting allowed by a set of rules in ZX-calculus. These core rules of ZX-calculus help in simulating the Clifford + T gate set easily. We can simulate simplified ZX-diagrams that don’t have an equivalent quantum circuit as ZX-diagram has the ability to flexibly portray quantum calculations than any other circuit [7]. The extraction of the circuits is also a direct procedure, where spiders are unfused and replaced with Clifford gates and CZ gates on the inputs and outputs as per requirement. 
These are not just mere flexible circuits but contain rich equational theory: the ZX-calculus and can be deformed arbitrarily. The core parameters of the ZX-calculus give a thorough theory for Clifford circuits, which can be efficiently classically simulated
\begin{figure}[h]
    \centering
    \includegraphics[width=9cm, height=5cm]{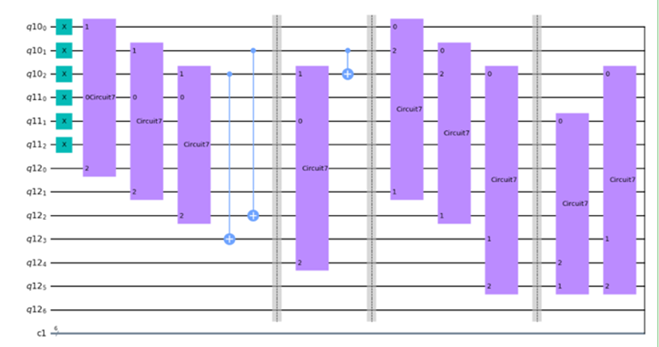}
\end{figure}
\begin{figure}[h !]
    \centering
    \includegraphics[width=9cm, height=5cm]{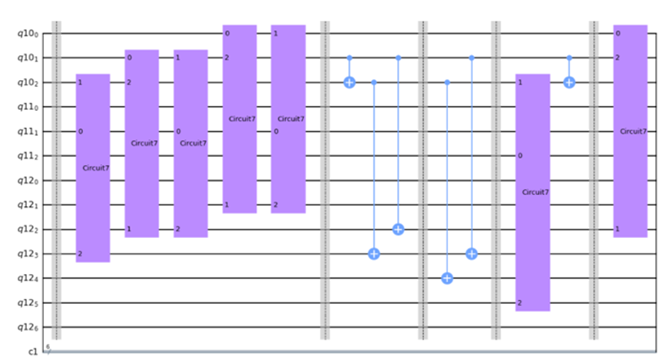}
\end{figure}
\begin{figure}[h !]
    \centering
    \includegraphics[width=9cm, height=5cm]{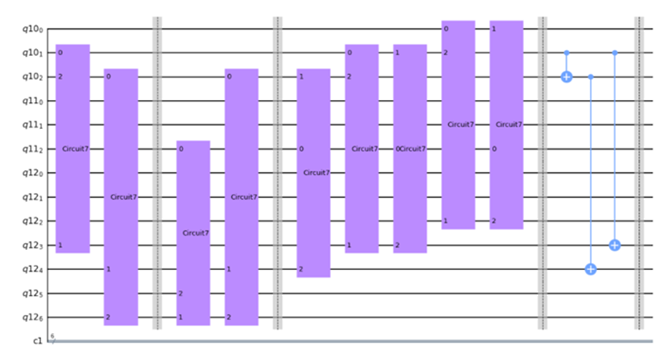}
    \caption{Integer Multiplier Circuit}
    \label{Fig 15}
\end{figure}.

Using ZX-diagrams, simplifications of ZX-diagrams can be derived with no quantum circuit analogue, since these diagrams are more flexible when compared to a quantum circuit representation. However, this flexibility comes with a disadvantage: By decomposing gates into smaller pieces quantum circuits can be interpreted as a ZX-diagram but the reverse is not valid. For effectively recovering a quantum circuit from any generalised ZX diagram, no specific method exists. Therefore a very crucial aspect of the  optimisation procedure is to retain enough information about the quantum circuit structure.

\section{Results}
In order to introduce the resistance to noise, in our integer multiplier circuit we have replaced the Toffoli gate with Clifford + T gate architecture. The circuit is then converted into a quantum gate using Python commands. The quantum gate is appended to the multiplier code replacing the Toffoli gate thereby making the circuit fault tolerant. The multiplier circuit using the Clifford gate is shown in Figure (15).

To perform the ZX calculus, we import an open-source Python library called “PyZX” in jupyter notebook. Firstly, we change the multiplier code into QASM code and transform it into basic quantum gates. The next step is to identify the gates present in the non-optimized multiplier circuit, followed by the conversion into a graph. The graph-like state is a crucial step before simplifying as it allows the optimization to result in a minimum number of internal spiders. The final step is the extraction of the optimized circuit and printing of the gates present in the circuit after optimization. In the end, we have verified the functionality of the circuit.

\begin{figure}[h!]
    \centering
    \includegraphics[width=9cm, height=8cm]{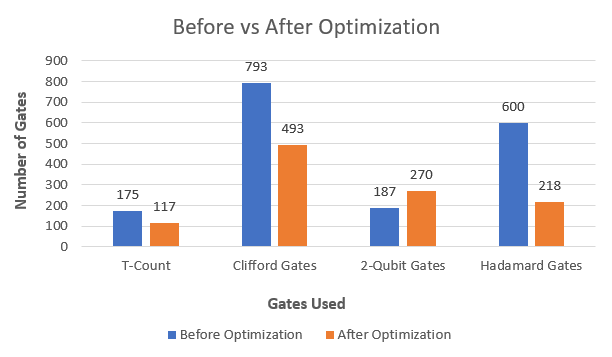}
    \caption{Optimization results}
    \label{Fig 16}
\end{figure}

For measuring the cost-effectiveness of our resultant optimized circuit, we perform a cost analysis in which T-gate count and ancilla inputs are the primary factors considered, the results of which have been tabulated in Table \ref{tab:my_label}. The T gate count is reduced from 742 to 488 after optimization which is shown in Fig \ref{Fig 16}. The Clifford gate count increases from 1044 to 2328. Where $n$ is the number of bits for which the adder is designed, in a conditional adder circuit, the T gate count is given by $21n + 14$, and the Toffoli gate array uses 7n T- gates. This brings the total T-count to \begin{equation}21{n}^{2}– 14\end{equation} The total ancilla inputs used in the design are given by \begin{equation}2n + 1\end{equation} For a 6-qubit circuit design, the T count is 742, with an ancilla  count of 13.
\begin{table}[!h]
\begin{center}
\begin{tabular} {||m{3cm}|m{2cm}|m{2cm}||} 
 \hline
 Features & Before \newline Optimization & After \newline Optimization\\ 
 \hline\hline
Number Of Qubits & 25 & 25\\ 
 \hline
 Number of Gates & 1786 & 2416\\
 \hline
 T-gate count & 742 & 488\\
 \hline
 Clifford Gate count & 1044 & 2328\\
 \hline
\end{tabular}
\end{center}
    \caption{Comparison of multiplier circuit before and after optimization for 6-bit multiplier}
    \label{tab:my_label}
\end{table}
\newline

\section{Conclusion}\label{conclusion}
In this work, a new method of optimizing quantum arithmetic circuits using ZX calculus is introduced. In particular, the fault-tolerant implementations of these circuits are considered and the trade-offs required with regard to the hardware cost are minimized by the use of this method. For the fault-tolerant implementation of the 6-qubit multiplier circuit, the optimization resulted in a 34\% reduction in the T-gate count with a trade-off of an increase in the Clifford gate count. The T-gate count is reduced when compared with the previous designs without any increase in the number of qubits which yielded an overall cost reduction for the hardware implementation of the circuit.

%\section{Acknowledgments}
%\noindent S.V. and S.G. have equally contributed to the project. Authors acknowledge the use of datasets from UCI Machine Learning Repository. Python programming language was used to carry out the entire reasearch.

\newpage
%\begin{thebibliography}{1}
\bibliographystyle{IEEEtran}
\bibliography{ref}% common bib file

\end{document}